\documentclass[final,5p,times]{amsci}

\usepackage{amssymb,amsfonts,amsmath,amstext,amsgen,amsopn,amsxtra,indentfirst,lscape}

\journal{Volume 103, Number 3, Pages 196--203 (May-June 2015 Issue)}

\begin{document}

\begin{frontmatter}

\title{The Next Great Exoplanet Hunt}

\author{Kevin Heng}
\ead[csh]{kevin.heng@csh.unibe.ch}
\address{University of Bern, Physics Institute, Center for Space and Habitability, Sidlerstrasse 5, CH-3012, Bern, Switzerland}
\author{Joshua Winn}
\ead[mit]{jwinn@mit.edu}
\address{Massachusetts Institute of Technology, Department of Physics, 77 Massachusetts Avenue, Cambridge, MA 02139-4307, United States of America}

\begin{abstract}
\textit{What strange new worlds will our next-generation telescopes find?\vspace{0.05in}\\
{\scriptsize Kevin Heng is an assistant professor of astrophysics at the University of Bern, Switzerland (Twitter: @KevinHeng1). He is a core science team member of the CHEOPS mission and is involved in the PLATO mission. Joshua Winn is associate professor of physics at the Massachusetts Institute of Technology, was a member of the Kepler science team, and is currently deputy science director of the TESS mission.} \vspace{0.05in}\\
Edited by Katie Burke.  \texttt{www.americanscientist.org }
} 
\end{abstract}

\end{frontmatter}

One of the most stunning scientific advances of our generation has been the discovery of planets around distant stars. Fewer than three decades ago, astronomers could only speculate on the probability for a star to host a system of exoplanets. Now we know the Universe is teeming with exoplanets, many of which have properties quite unlike the planets of our Solar System. This has given birth to the new field of exoplanetary science, one of the most active areas of astronomy, pursued by many research groups around the world.

Although there are many ways of detecting exoplanets, the most successful in recent years has been the transit method, in which the exoplanet reveals itself by transiting (passing directly in front of) its host star, causing a miniature eclipse. These eclipses are detected by telescopes that are capable of precisely tracking the brightnesses of many stars at the same time. Since its launch in 2009, NASA's Kepler space telescope used this technique to resounding effect, finding more than 1,000 confirmed transiting exoplanets. The success of the Kepler mission has inspired the next generation of exoplanet transit-hunting machines: a fleet of new space telescopes and complementary ground-based telescopes. They will expand our inventory of strange new worlds and further our quest to find Earth-like exoplanets and search them for signs of life.

The long-term strategy of exoplanet hunting is easy to state: find exoplanets, characterize their atmospheres and, ultimately, search for chemical signatures of life (biomarkers). These efforts strive to answer questions such as: Which molecules are the most abundant? What is the typical cloud coverage, temperature and wind speed? Is there a solid surface beneath all the gas? The answers to these questions help establish whether an exoplanet is potentially habitable. And unless we can detect radio broadcasts from an intelligent civilization or build starships to visit the exoplanet itself, these atmospheric characterizations seem the most promising---and to some, the only---way of discerning whether a given exoplanet is actually inhabited. Molecular oxygen, or ozone, for example, would be circumstantial, but not definitive, evidence for life, because there are plausible ways of producing it abiotically using the known laws of physics and chemistry. Subsurface life may exist on an exoplanet, but we probably have no chance of detecting it with astronomical observations.

This logical sequence of exploration is largely mirrored in the series of space missions that NASA and the European Space Agency (ESA) have planned, approved, and constructed---or will construct. Exoplanet detection is poised to make the transition from cottage industry to big science, while atmospheric characterization is inexorably finding its feet.

\vspace{0.15in}
\noindent
\textbf{Bright, Twinkling Stars}
\vspace{0.05in}

Early efforts to detect transiting exoplanets in the early 2000s used ground-based telescopes, which face daunting obstacles. One obstacle is the annoying tendency of the Sun to rise every morning. This foils the attempt to monitor stars as continuously as possible. Transits that occur during daytime are missed. Another obstacle is the Earth's atmosphere, a turbulent and constantly varying screen that causes apparent fluctuations in starlight much stronger than the transit signals we seek to detect. Even on a seemingly clear night, the atmosphere alters the passage of starlight in a manner that depends on both wavelength and brightness, through the phenomena of extinction and scintillation. Extinction is why the setting sun is red and dim; scintillation is why stars twinkle. To correct for these effects, astronomers point their telescopes at groups of stars with similar colors and brightness, trusting that the atmosphere will affect them all equally---whereas a transiting exoplanet would cause only one of them to temporarily dim. Using these reference stars, exoplanet transits can be flagged and confirmed.

From our vantage point on Earth, bright stars are rarer than faint ones, simply because of geometry---casting our net farther into space would create a celestial sphere that encompasses more stars, which tend to be fainter on average because of the greater distances involved. The brightest stars are distributed uniformly across the entire sky and are therefore widely separated in angle. Finding reference stars to confirm an exoplanet transit becomes harder for bright stars: the only comparable reference stars may be located so far away on the sky that one can no longer assume they are affected equally by the atmosphere. This helps to explain the otherwise paradoxical fact that the ``best and brightest" few thousand stars in the sky are relatively unexplored for transiting exoplanets.

\begin{figure}
\begin{center}
\includegraphics[width=\columnwidth]{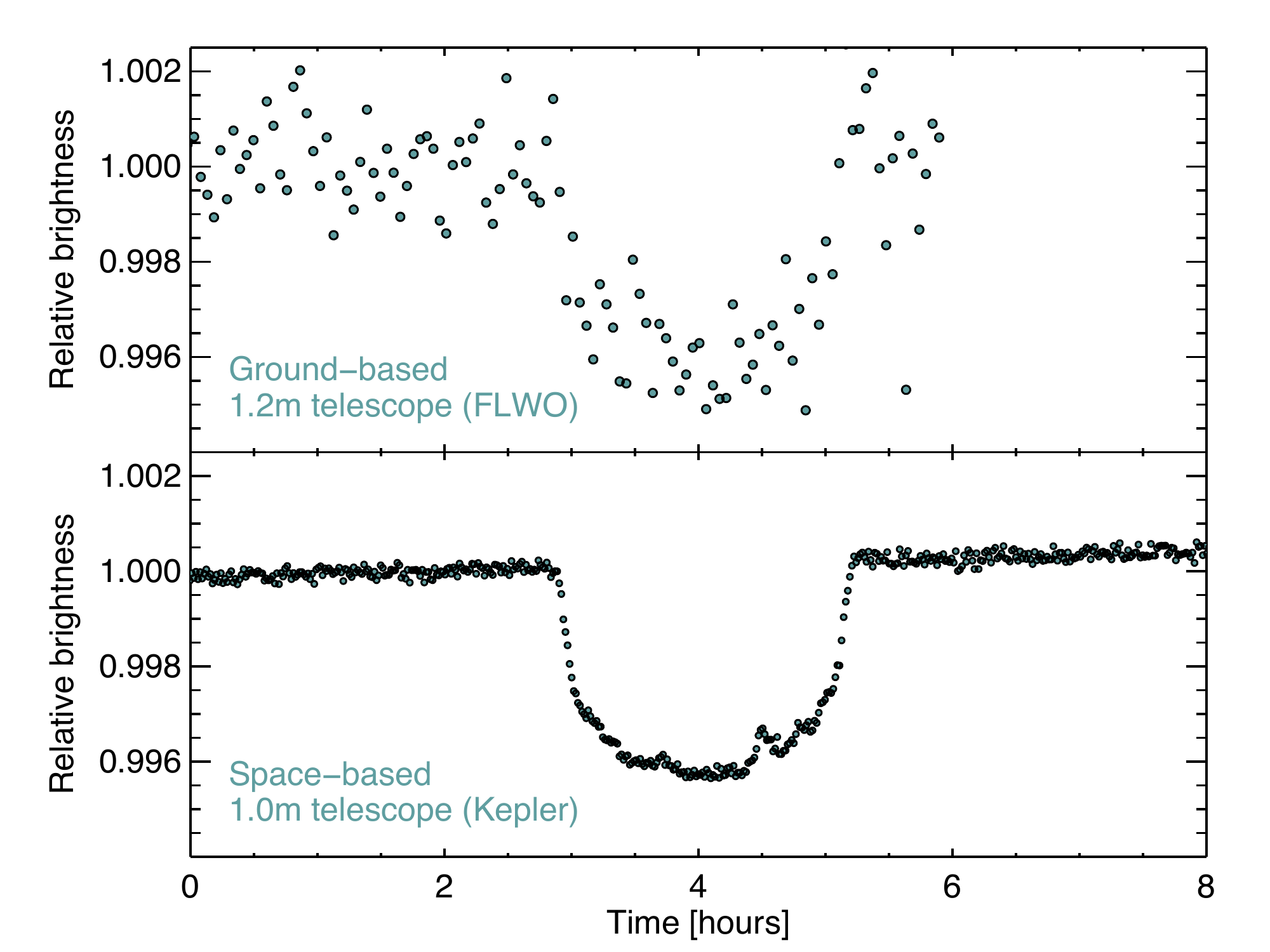}
\end{center}
\caption{\textbf{Why we need space telescopes to observe transiting exoplanets.} The top panel shows a time series of brightness measurements of the exoplanet HAT-P-11b, using an Arizona-based telescope with a 1.2-meter diameter. The dip of 0.4\%, lasting a few hours, is due to the miniature eclipse of the star by an exoplanet a shade larger than Neptune. The large scatter in the measurements is due to failure to correct adequately for atmospheric effects and the interruption just after the transit is due to morning twilight. The bottom panel shows an observation of the same exoplanet with the 1.0-meter Kepler space telescope. Despite being a smaller telescope, the absence of atmospheric effects and the uninterrupted observations enable a vastly improved view of the transit. It is even possible to tell that the exoplanet crossed over a dark spot on the surface of the star (known as a ``star spot"), based on the slight upward flicker that was seen in the second half of the transit (at a time coordinate of about 4.5 hours).}
\end{figure}

Astronomers realized that these difficulties may be circumvented by launching telescopes into space. Although he was not specifically thinking about exoplanet transits, the late Princeton astronomer Lyman Spitzer lobbied tirelessly in the 1960s and 1970s for what would later become known as the Hubble Space Telescope. Placed above Earth's atmosphere eliminates some of the obstacles to precise measurements. For exoplanet hunters, there is no need to find bright reference stars and the precision is limited only by photon-counting noise (sometimes termed ``shot noise")---the inevitable fluctuations in the signal caused by the discrete and random nature of photon emission. 

Even with a space telescope, detecting transiting exoplanets is a fight against long odds. The vast majority of exoplanets do not transit. The exoplanet's orbit must be aligned nearly edge-on, as we see it from Earth, in order for transits to occur. The probability for transits is equal to the stellar diameter divided by the orbital diameter, which is only 0.1\% for an Earth-like orbit around a Sun-like star. For this reason, a meaningful transit survey must include tens of thousands of stars, or more. Because faint stars far outnumber bright ones in any given region of the sky, a practical strategy is to monitor a rich field of relatively faint stars. This is precisely what the Kepler space telescope did, staring at about 150,000 stars in a small, 115-square-degree patch of sky in the constellations of Cygnus and Lyra. Unfortunately, most of the Kepler discoveries involve stars that are too faint for the atmospheric characterization of their exoplanets.

Transiting exoplanets offer the potential for atmospheric characterization. During a transit, a small portion of the starlight is filtered through the exoplanet's atmosphere. By comparing the spectrum of starlight during a transit to the starlight just before or after the transit, we can identify absorption features due to atoms and molecules in the exoplanet's atmosphere. Another approach is to measure the light emitted from the exoplanet, by detecting the drop in brightness when the exoplanet is hidden behind the star, an event known as an ``occultation". 

The practical value of transits and occultations depends critically on the brightness of the star. The higher the rate of photons (particles of light) that a star delivers to Earth, the faster we can accumulate information about its exoplanets. This is because many astronomical observations are limited by photon-counting noise. The greater the number of photons we collect, the smaller the effects of shot noise and the higher the signal-to-noise ratio. This in turn allows us to search for smaller exoplanets, which produce smaller transit signals. To put the smallness in perspective, an Earth-sized exoplanet transiting a Sun-like star produces a total dimming of only 84 parts per million and its atmospheric signal would be smaller by at least an order of magnitude. To measure the spectrum of an exoplanetary atmosphere, one needs to divide up the starlight according to wavelength. In essence, a small signal must be split into even smaller ones, an endeavour only feasible if the signal-to-noise ratio is very high at the outset. And this is only possible for bright stars. The Kepler stars are typically more than a million times fainter than the brightest naked-eye stars such as Sirius, Vega and Alpha Centauri. 

These challenges are easier to overcome when the exoplanet is large, is located close to its star and has a ``light" (low mean molecular mass) or nonmetallic atmosphere. This is why hot Jupiters, hydrogen-dominated gas giants orbiting their stars closer than Mercury is to our Sun, were the first to have their atmospheres characterized by astronomers.

To appreciate how demanding the observations can be, consider the transiting super-Earth known as GJ 1214b, the most favorable system for atmospheric characterization among all the known transiting exoplanets smaller than Neptune, because it is located relatively close to us and resides in a 1.6-day orbit. Even in this most favorable case, characterizing GJ 1214b's atmosphere required observing with the Hubble Space Telescope for a record-breaking 60 orbits, which translates roughly into a cost of about \$12 million. (Observing time on the Hubble Space Telescope is quantified by the number of 90-minute orbits of the spacecraft around the Earth and is awarded competitively. Most successful proposals request only a few orbits.) Unfortunately, the transmission spectrum recorded did not achieve a definitive conclusion: Does GJ 1214b have a ``heavy" (metallic) atmosphere? Or does it have a hydrogen-dominated atmosphere enshrouded by clouds? To scale up this endeavor to thousands of exoplanetary atmospheres, it would make sense to target the brightest possible stars.

Combining the Kepler discoveries with those made by other groups, astronomers now know of about 2,000 confirmed exoplanets and another 4,000 exoplanet candidates around other stars. These impressive-sounding statistics obscure the fact that the number of exoplanets for which we can meaningfully characterize the atmosphere is quite small---about a dozen. And they are mainly gas-giant exoplanets; most of them are significantly larger than Earth and none of them are likely to be habitable.

\begin{figure}
\begin{center}
\includegraphics[width=\columnwidth]{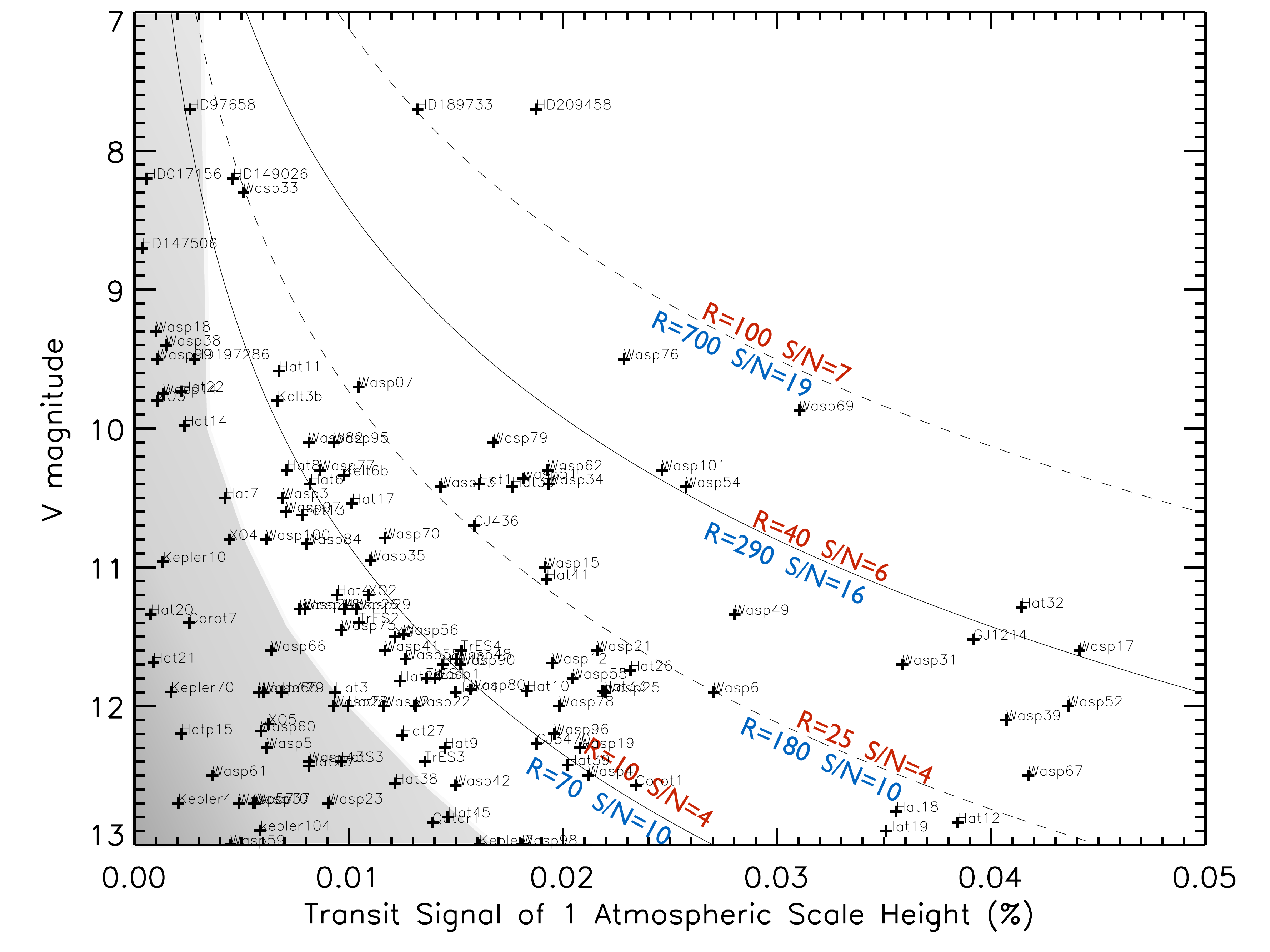}
\end{center}
\caption{\textbf{Why exoplanets are not all equal when it comes to atmospheric characterization.} The figure shows the brightness of star versus the ease at which the atmosphere of the exoplanet may be characterized. The stellar brightnesses are recorded in the visible range of wavelengths (hence the ``V" label). Lower V magnitudes correspond to brighter stars. A typical length scale in an atmosphere is the pressure scale height, the distance over which the density of an atmosphere changes substantially. Generally, exoplanets with higher temperatures, less metallic atmospheres and weaker gravities tend to have larger scale heights---the atmospheres are thus puffier and easier to detect and characterize. The various curves indicate the regions of the graph where the signal strength (signal-to-noise ratio or S/N) and the resolution of the spectrum of the exoplanetary atmosphere obtained are of the stated values. The numbers in red and blue are for the Hubble Space Telescope and James Webb Space Telescope, respectively. Exoplanetary atmospheres in the gray shaded region are unobservable due to low S/N. Courtesy of David Sing (University of Exeter).}
\end{figure}

For astronomers, the next step is crystal clear: detect transiting exoplanets as small as the Earth around the brightest and nearest stars, which would give us the best chance of characterizing their atmospheres and searching for biosignature gases---ones that can be uniquely attributed to life and are present in enough abundance that our remote sensing techniques may robustly detect them.
 
\vspace{0.15in}
\noindent
\textbf{Exoplanet-Hunting Machines}
\vspace{0.05in}

After the failure of two of its reaction wheels (which are responsible for maintaining the telescope's orientation in space), the Kepler mission is unable to keep staring at the same field of stars. The telescope and camera are still functioning but they are unable to hold a steady pointing except if they are aimed close to the ecliptic, the plane of Earth's orbit. The mission has recently been re-christened ``K2" and is now surveying various star fields in the ecliptic. This will turn up more transiting exoplanets around bright stars, but leaves most of the sky to be explored by its successors.

In 2012, ESA selected the Characterising Exoplanet Satellite (CHEOPS) as their first small-class space mission (at a cost of about 100 million euros). Led by the Swiss astrophysicist Willy Benz, CHEOPS builds on the Swiss heritage of exoplanet hunting. Widely recognised for having found the first exoplanet orbiting a Sun-like star, Geneva Observatory continues to lead the world at building the most precise spectrographs for the Doppler detection of exoplanets: measuring the to-and-fro wobble of stars as they are gravitationally pulled around by their exoplanets. Doppler data allow the masses of the exoplanets to be measured.

Shortly afterwards, NASA approved the Transiting Exoplanet Survey Satellite (TESS), building on the heritage of the Kepler mission, with a cost of about \$200 million and a launch scheduled for 2017, the same year as CHEOPS. Barely a year later, the exoplanet-hunting zeitgeist was further recognized by ESA, which approved the Planetary Transits and Oscillations of Stars (PLATO) mission---yet another transit detection machine---to be launched in 2024 at a price tag of half a billion euros. Each of these machines aims to take the next logical steps after Kepler and discover transiting exoplanets around bright stars. 

\begin{figure}
\begin{center}
\includegraphics[width=\columnwidth]{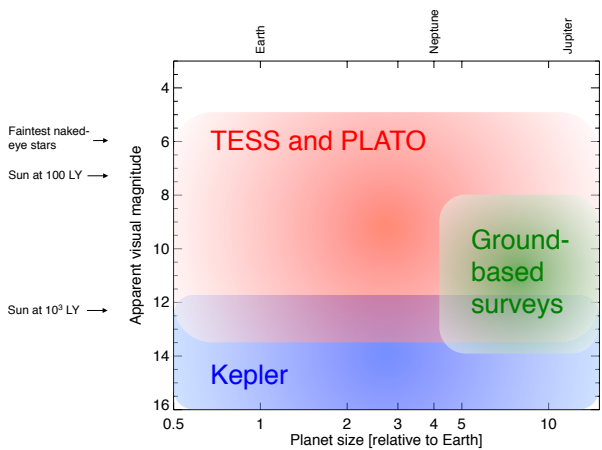}
\end{center}
\caption{\textbf{Brightness of stars that host transiting exoplanets.}  The space missions Kepler, TESS and PLATO are sensitive to a wide range of exoplanet sizes from sub-Earth to Neptune- and Jupiter-like, whereas ground-based surveys are capable of detecting objects that are mostly Neptune-sized or larger.  The orbital periods of the exoplanets are not shown in the plot, but it should be noted that TESS is mainly sensitive to short-period (weeks) exoplanets, while PLATO will discover long-period (months to years) ones.  The MEarth survey is not depicted in this schematic, because it operates at far-red, rather than visible wavelengths; despite being a ground-based survey, it can detect exoplanets smaller than Neptune, because it focuses exclusively on small stars, which radiate mainly in the far-red and infrared wavelengths.  ``LY" means one light year.}
\end{figure}

Why do we need so many different exoplanet transit space missions? Leaving geopolitical considerations aside, they are all different beasts with different hunting strategies. TESS will pursue the strategy of scanning the entire sky in a systematic manner, concentrating on each sector of the sky for about 27 days. It will use four optical cameras, each with an unusually wide (24 by 24 degrees) field of view, to monitor about 200,000 stars. Since the rule of thumb for confirming a detection is to record three transits---so as to establish periodicity---this largely restricts TESS to detecting exoplanets on short, 9-day orbits, except for two continuous viewing zones where it will monitor the sky for up to an entire year. Centered on the ecliptic poles, these viewing zones are chosen to be compatible with the ease of using the James Webb Space Telescope (JWST) to follow up on TESS targets. TESS is a discovery mission in the truest sense, in that it will search for previously unknown exoplanet transits. The TESS team expects to find about a thousand transiting exoplanets, including hundreds that are smaller than Neptune and a few dozen Earth-sized ones. 

By contrast, CHEOPS uses a narrow-field camera (0.3 by 0.3 degrees) to examine one star at a time. It was originally conceived to search for transits in systems where Doppler measurements by the Geneva exoplanet group have previously detected the presence of exoplanets, but for which it is unknown whether the exoplanet's orbit is oriented correctly for transits. Exoplanets with both transit and radial velocity measurements are golden targets for atmospheric characterization, since their masses, radii and surface gravities would be known, which allow for a more decisive interpretation of their atmospheric properties. CHEOPS can also perform a useful role by staring at systems already known to transit (from TESS or other surveys), confirming signals and giving better estimates of exoplanet sizes and orbital periods.

The most important characteristic of a telescope is the size of its mirror and therefore its light-gathering power. Mirror size is always a compromise between scientific goals, engineering considerations and cost. Bigger mirrors collect more photons, enabling more precise brightness measurements and the detection of smaller transiting exoplanets---but bigger mirrors usually have smaller fields of view and are more complex and costly. TESS uses four cameras with mirrors that are only 10 centimeters in diameter, offering relatively low cost and wide fields of view. CHEOPS will deploy a 30-centimeter mirror, allowing nine times as much light to be collected every second. All other things being equal, this means a single transit observed by CHEOPS bears as much information as nine repeated observations by TESS. The practical challenge for CHEOPS is to examine a large enough sample size of stars with exoplanets, since only a fraction of them will---by chance---transit their stars. 

Conversely, a major challenge for TESS is to coordinate the ground-based telescope observations that are needed to detect the Doppler signals and measure their masses. It turns out that the ease or difficulty of performing these measurements is related to the type of star involved. Astronomers categorize stars based on their surface temperatures, which are revealed by their spectra. They are labeled in the following order of decreasing temperature: O, B, A, F, G, K, M (which inspired the mnemonic, ``Oh, be a fine girl/guy, kiss me."). O stars are early-type stars and live very short lives---only millions of years---compared to their later, Sun-like brethren (G stars), which have lifetimes of billions of years. Early-type stars also tend to be large, implying that it would be harder to detect the transits of small exoplanets around them. M stars are late-type stars and are the longest-lived, potentially lasting up to the age of the Universe. 

Stars of spectral type F and earlier (meaning A, B and O stars) rotate more rapidly, which broaden the spectral lines of atoms or molecules in their atmospheres. Since making radial velocity measurements depends on being able to precisely pin down the positions of these spectral lines associated with the star, broader lines are an impediment. Consequently, it becomes more challenging to measure the mass of the exoplanet. It is for this reason that TESS has chosen to focus on stars from mid-F and mid-M. Small, late-type stars are our best chances, in the short term, of detecting and characterizing an Earth-like exoplanet. This is partly because the transit of a given exoplanet produces a bigger signal when it is around a smaller star: The loss of light during a transit is equal to the square of the size ratio of the exoplanet and star. Another advantage of small stars is that they are much less luminous, implying that exoplanets with the same surface temperature as Earth would reside in closer-in, shorter-period orbits---making them much easier to detect. Radial velocity measurements are also easier for bright stars, as one is again performing spectroscopy. The faintness of the Kepler stars explains why only a handful of them have radial velocity measurements. 

Doppler measurements are important for another reason---by pinning down the mass of the transiting object, one can definitively identify it as an exoplanet. The vastness of the Universe and the creativity of Nature produce a variety of astrophysical ``false positives": astrophysical signals that mimic transits but are caused by stars and brown dwarfs. Brown dwarfs are objects intermediate in mass between exoplanets and stars---between 13 and 80 Jovian masses---but often have radii similar to that of Jupiter. A transiting brown dwarf is therefore impossible to distinguish from a transiting gas-giant exoplanet, unless its mass can be measured. Exoplanetary transits can also be mimicked by grazing stellar binaries---a pair of stars residing in an orbit that is not quite edge-on to the observer so that each star barely obscures the other during conjunction. This near-obscuration produces a tiny eclipse that can appear similar to an exoplanet transit. Alternatively, the light from a pair of eclipsing stars can be blended together with the constant light from a star that is nearby on the sky, causing the signal of the stellar eclipses to be diluted down to exoplanet-sized proportions. To complicate matters, there are also ``false alarms", spurious signals caused by our imperfect understanding of the light detectors or by stellar activity. Stars are hardly perfect light bulbs---sometimes, they flicker and erupt, a fact that was unthinkable up till the time of Galileo.  The ability to thoroughly understand the false-positive occurrence rate in one's survey or mission may ultimately be the difference between failure and success.

Ground-based transit surveys are also advancing rapidly. They are less capable than spacecraft, but also much cheaper and play a complementary role. Building on the Wide Angle Search for Planets (WASP) survey, the UK-led Next-Generation Transit Survey (NGTS) is an array of fully robotic telescopes sitting on a desert plateau in Chile and searching for transits around K and M stars. The MEarth project, led by David Charbonneau of Harvard University, discovered the super Earth GJ 1214b and continues to search for small exoplanets around M stars. The SPECULOOS project, led by the Belgian astronomer Micha\"{e}l Gillon, is also targeting late M stars---the dimmest, smallest and longest-lived of its category. In addition to dealing with the Earth's atmosphere, ground-based astronomers have to worry about the complications associated with M stars, including enhanced flaring and stellar activity, as well as star spots---imperfections on their surfaces associated with magnetic fields---that sometimes rival the size of any orbiting exoplanet. These challenges are offset by the fact that M stars are abundant---they make up about three-quarters of the stars in our solar neighborhood. 

This fleet of space missions, complemented by intensive efforts from the ground, will lay the foundation for building up a target list that may finally allow atmospheric characterization of exoplanets to be statistical in nature.

\vspace{0.15in}
\noindent
\textbf{Awaiting the Behemoths}
\vspace{0.05in}

This grand mobilization of exoplanet detection machines awaits the launch of the much-anticipated James Webb Space Telescope (JWST) in 2018. More than a decade in the making, the JWST is the successor to the Hubble Space Telescope, but with ``eyes" optimized for infrared observations rather than visible and ultraviolet light. Capped at a cost of \$8 billion, it is humankind's most expensive telescope. It is a behemoth with a 6.5-meter mirror, armed with a suite of cutting-edge instruments capable of taking spectra from the faintest and most distant objects in the sky. Recognizing the importance of the JWST for exoplanet science, NASA recently tweaked the design of one of its instruments to enable better measurements of exoplanet transits around bright stars.

The catch is that the JWST is a general observatory built for multiple branches of astronomy, including cosmology, black hole physics, galaxy formation and evolution, protoplanetary disks, exoplanets, the interstellar medium, the first stars, and so on. The mission lifetime of the JWST (5.5 to 10 years) is limited by the supply of hydrazine fuel needed to maintain the spacecraft's orbit. Coupled with the constraint that its observing time has to be shared with the other, non-exoplanet astronomers, this means that the JWST will be restricted to characterizing a small number of exoplanetary atmospheres. While there is currently no shortage of intriguing exoplanetary atmospheres for the JWST to characterize, the hope is that we will have found several Earth-like exoplanets orbiting bright stars by the time of its launch. In an ideal world, we would want an Earth analogue orbiting a M star located sufficiently close to Earth that there is a chance of detecting the presence of biosignature gases in its atmosphere (assuming life exists on that exoplanet). Statistically speaking, such targets are likely to exist, but none have yet been identified.

The PLATO mission of ESA, scheduled for launch in 2024, will be getting a much later start than TESS and CHEOPS at finding targets for JWST. However, PLATO is the most ambitious mission of the three, aiming to build a catalogue of true Earth analogs---Earth-like exoplanets orbiting within the habitable zone of Sun-like stars. Inspired by the WASP ground-based survey, it will feature an array of 34 telescopes with 12-centimeter mirrors. Its observing strategy borrows from that of Kepler, CHEOPS and TESS: a sweep of the entire sky with the possibility of ``step-and-stare" phases lasting two to five months. Each sector of the sky is observed for two to three years, sufficiently long to uncover exoplanets on year-long orbits. The PLATO team will target and monitor a million stars and expects to detect over 1,000 Earths and super Earths on various orbits, complete with follow-up Doppler measurements using next-generation spectrographs from the ground. PLATO will truly carry the categorization of the bulk properties of exoplanets to the next level. 

What will follow the transit discovery missions? During the competitive selection process won by CHEOPS and TESS on either side of the Atlantic, there were two proposed missions for telescopes specializing entirely in atmospheric characterization, named EChO and FINESSE. These fell by the wayside, but will likely be revived with an even stronger motivation after all of the best and brightest targets have been discovered. Such a mission would need to record spectra from the visible to the mid-infrared range of wavelengths. Information from the visible wavelengths would encode a combination of how reflective the atmosphere is, tell us whether it is cloudy and identify the inert gas responsible for most of its mass. It would also identify atoms such as sodium or potassium, which have previously been detected in several hot Jovian atmospheres. Identifying the presence of molecules such as water, ammonia, carbon monoxide, carbon dioxide and methane would require an infrared spectrum. By building up a census of the chemical compositions of exoplanetary atmospheres in our cosmic neighborhood, we may begin to address questions like whether ``water worlds"---ocean-dominated exoplanets---are common in the Universe.

Some astronomers are also investigating cheaper, ground-based alternatives to building space telescopes. The Dutch astronomer Ignas Snellen has suggested the idea of constructing ``light buckets"---arrays of small- to medium-sized telescopes for recording exoplanet transits. Most astronomical telescopes are designed to capture sharp images of distant objects. But for the purpose of measuring a transit, image sharpness is largely irrelevant---all one needs is to collect enough light to register a robust signal of the rise and ebb of starlight. In principle, the same amount of light may be collected by a large array of hundreds of telescopes working together to mimic the light-collecting area of a 40-meter (or larger) telescope---at a fraction of the cost.

Other astronomers are chasing the holy grail of taking a photograph of a second Earth, a technique known in exoplanetary science as ``direct imaging". The key challenge is to block out the starlight, which is a million to a billion times more brilliant than the light of the exoplanet (depending on whether one is operating in the visible or infrared range of wavelengths). There are two schools of thought on how to achieve this obscuration: either to install an internal coronagraph as part of the design of the space telescope; or to fly a separate star shade that positions itself far away from the telescope. Flying a star shade shifts the engineering complexities away from the telescope, implying that a run-of-the-mill space telescope may be used---but at the price of having two separate spacecraft. Several variations of these ideas have been considered for decades, with the most recent study of the star shade being led by Sara Seager at MIT. 

More audaciously, perhaps the process of mapping the chemical and physical diversity of the exoplanetary atmospheres in our cosmic backyard can be accelerated via the private sector. Space flight is inexorably being commercialized. There is even a private enterprise (Mars One) aiming to send humans to colonize Mars. If the same resources were directed towards building a dedicated observatory for characterizing exoplanetary atmospheres, perhaps we could complete our census within a few decades. One idea would be to hold the X Prize equivalent of a competition for constructing such an observatory, fueled by private investors. Mapping out the habitable real estate of our cosmic neighborhood seems at least as exciting as visiting the soils of our Solar System.

Amidst this rush of activity, we often have to explain ourselves to our fellow physicists, who perceive exoplanetary science as being ``not fundamental" and qualifying only as ``applied physics." But purity does not always equate with importance, and what is not pure is not always trivial. Although exoplanet hunters are certainly not expecting to stumble upon new insights into grand unified theories, the scientific stakes are no less high: to remove humanity from the center of the physical, chemical and biological universe, thus completing the Copernican-Galilean revolution, which established that the Earth revolves around the Sun. To quote the American exoplanet hunter Sara Seager: ``Hundreds or a thousand years from now, people embarking on interstellar travel will look back and remember us as the society that first found the Earth-like worlds." The hunt is on.

\label{lastpage}

\end{document}